\documentclass[letterpaper, 10 pt, conference]{ieeeconf} 
\IEEEoverridecommandlockouts 

\usepackage{amsmath,amssymb,amsfonts}
\usepackage{bbold}
\usepackage[utf8]{inputenc}
\usepackage{bbm}
\usepackage{makecell}
\usepackage{cite}
\usepackage{subfigure}
\usepackage{placeins}
\usepackage{babel,blindtext}
\usepackage{verbatim}
\usepackage{hyperref}
\usepackage{xurl}
\usepackage{cleveref}
 
\usepackage{cite}
\usepackage{algorithmic}
\usepackage[graphicx]{realboxes}
\usepackage{textcomp}
\usepackage[table,xcdraw]{xcolor}
\usepackage{multirow}
\usepackage{array}
\usepackage{tikz}
\usepackage{siunitx}
\usepackage{lscape}
\usepackage{adjustbox}
\usepackage{tabularx}
\usepackage{mleftright}
\usepackage{etoolbox}
\usepackage{xcolor}
\usepackage{booktabs}

\title{\LARGE \bf
GC-TTS: Few-shot Speaker Adaptation with Geometric Constraints
}

\author{Ji-Hoon Kim$^{1}$, Sang-Hoon Lee$^{2}$, Ji-Hyun Lee$^{1}$, Hong-Gyu Jung$^{2}$, and Seong-Whan Lee$^{1}$
\thanks{This work was supported by Institute for Information \& communications Technology Planning \& evaluation (IITP) grant funded by the Korea government (MSIT) (No. 2019-0-00079, Artificial Intelligence Graduate School Program (Korea University)) and the Netmarble AI Center.}
\thanks{$^{1}$J.-H. Kim, J.-H. Lee, and S.-W. Lee are with the Department of Artificial Intelligence, Korea University, Anam-dong, Seongbuk-ku, Seoul 02841, Korea.
        {\tt\small \{jihoon\_kim, jihyun-lee, sw.lee\}@korea.ac.kr}}
\thanks{$^{2}$S.-H. Lee and H.-G. Jung are with the Department of Brain and Cognitive Engineering, Korea University, Anam-dong, Seongbuk-ku, Seoul 02841, Korea.
        {\tt\small \{sh\_lee, hkjung00\}@korea.ac.kr}}
        }

\begin{document}

\maketitle
\thispagestyle{empty}
\pagestyle{empty}

\begin{abstract}
Few-shot speaker adaptation is a specific Text-to-Speech (TTS) system that aims to reproduce a novel speaker's voice with a few training data. While numerous attempts have been made to the few-shot speaker adaptation system, there is still a gap in terms of speaker similarity to the target speaker depending on the amount of data. To bridge the gap, we propose GC-TTS which achieves high-quality speaker adaptation with significantly improved speaker similarity. Specifically, we leverage two geometric constraints to learn discriminative speaker representations. Here, a TTS model is pre-trained for base speakers with a sufficient amount of data, and then fine-tuned for novel speakers on a few minutes of data with two geometric constraints. Two geometric constraints enable the model to extract discriminative speaker embeddings from limited data, which leads to the synthesis of intelligible speech. We discuss and verify the effectiveness of GC-TTS by comparing it with popular and essential methods. The experimental results demonstrate that GC-TTS generates high-quality speech from only a few minutes of training data, outperforming standard techniques in terms of speaker similarity to the target speaker.
\end{abstract}

\begin{keywords}
speech synthesis, few-shot learning, speaker adaptation, fine-tuning, speaker embedding
\end{keywords}

\begin{figure*}[!]
  \centering
  \scriptsize
  \centerline{\includegraphics[width=\textwidth]{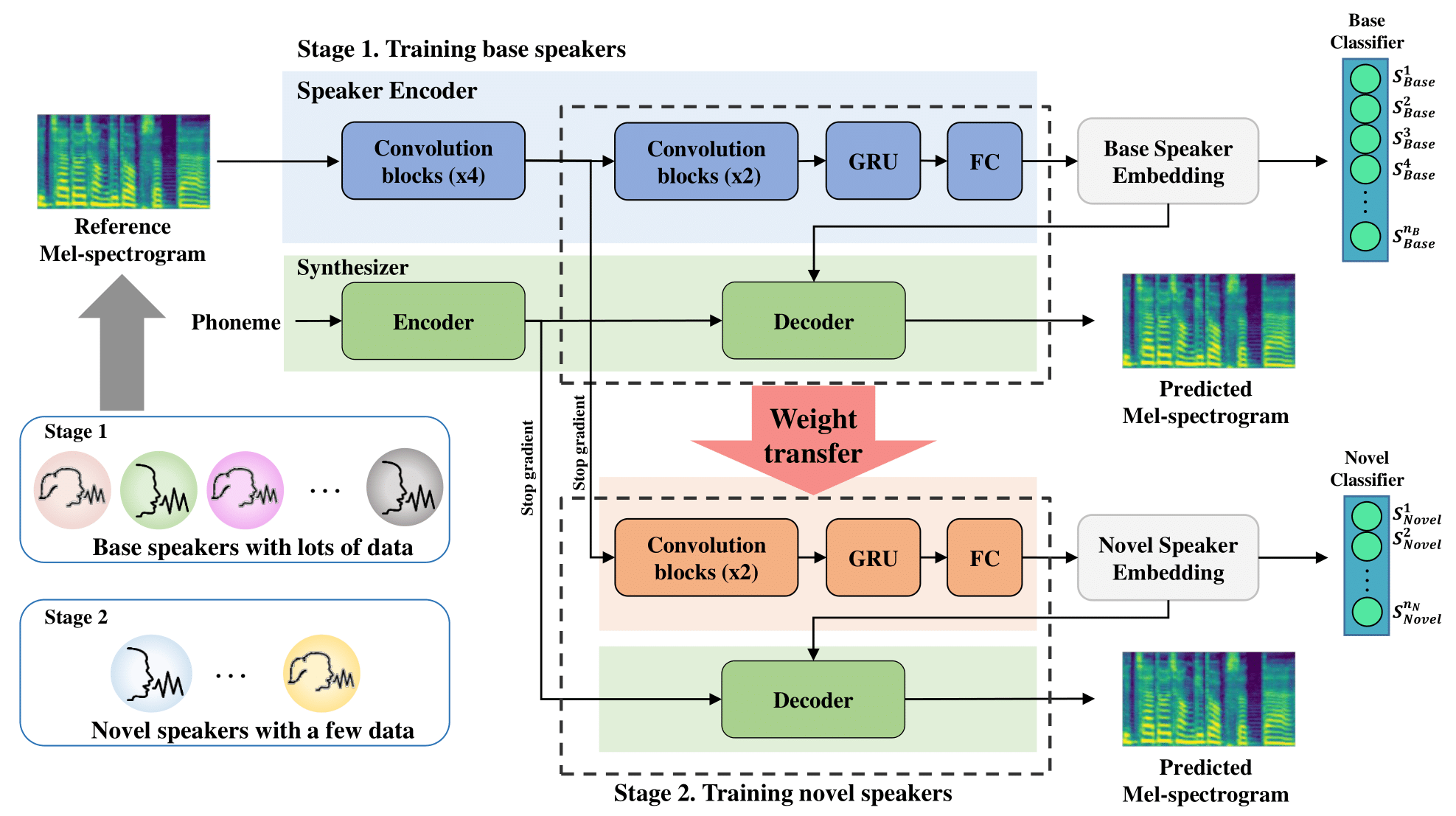}}
  \caption{Network architecture. In training stage 1, we pre-train the network for base speakers with a large amount of data. In training stage 2, we first freeze the weights of the Tacotron2 encoder, attention module, and four bottom convolution blocks in the speaker encoder. Then, we replicate the other modules which are shown using broken lines. Finally, we build a classifier for novel speakers. $n_B$ and $n_N$ refer to the number of the base and novel speakers, respectively. }
  \label{fig:network}
\end{figure*} 

\section{INTRODUCTION}
\label{sec:intro}
Text-to-Speech (TTS) system aims to convert a normal text into human speech. Conventional 
TTS pipelines are based on a number of complex stages, each of which requires laborious feature engineering such as text normalization\cite{hunt1996unit,black2007statistical}. After the emergency of Tacotron\cite{wang2017tacotron} and DeepVoice\cite{arik2017deep}, TTS model can be simplified and generate natural speech from text inputs in an end-to-end manner. While recent TTS models achieve the synthesis of intelligible speech in an end-to-end manner, training these system requires tons of high-quality data collected in controlled settings\cite{chen2015deep,shen2018natural:1,li2019neural,prasetio2019deep,donahue2020end,ren2020fastspeech,lee2020multi}. Given this high cost, there have been increasing demands for a few-shot speaker adaptation system which is a technique to function with a few amount of data. 

Existing studies for a few-shot speaker adaptation can be categorized into two categories. One is to utilize an independent neural encoder to extract speaker embeddings which are then combined with linguistic features of a pre-trained TTS model \cite{li2017deep,jia2018transfer:7,cooper2020zero}. While these approaches lead to a simple and efficient adaptation, they result in poor adaptation quality, especially in speaker similarity to the target speaker. Another is to fine-tune a pre-trained TTS model to novel speakers with a small amount of adaptation data\cite{arik2018neural:5, chen2018sample:6,inoue2020semi:7, moss2020boffin,chen2020ada}. They outperform the former one in terms of speaker similarity. However, the similarity is still lower than when the model is trained with a large amount of training data. Moreover, it is prone to destroy the speaker embedding space well trained for base speakers\cite{mao2015learning,kim2020few}, resulting in degraded generation performance for base speakers.

In this paper, we propose GC-TTS, a few-shot speaker adaptive TTS framework which achieves high-quality speaker adaptation for novel speakers while preserving the ability to synthesize natural speech for base speakers. To achieve this, GC-TTS leverages two geometric constraints\cite{jung2020few:11} which shows impressive results in few-shot image classification. GC-TTS is trained in two stages. In training stage 1, we first train the model for base speakers with a sufficient amount of data (i.e. $10$ hours or more). In training stage 2, we continue to train the model for novel speakers on a few minutes of data with two geometric constraints. The first constraint, when used with cross-entropy loss, enables speaker embeddings of utterances from the same speaker to have a small angular distance. The second constraint makes those of utterances from different speakers have large angular distances in the embedding space. Moreover, we only update parts of network parameters in training stage 2 so that we can prevent overfitting and fully utilize the prior knowledge in an effective manner. 

While prior studies have only focused on generating intelligible speech for novel speakers, to the best of our knowledge, this is the first attempt to adapt a TTS model to novel speakers without diminishing generation performance for base speakers. The experimental results indicate that GC-TTS is capable of synthesizing improved speech over standard techniques, especially in speaker similarity to the target speaker. In addition, visualization of the speaker embedding space shows GC-TTS constructs discriminative speaker embeddings of novel speakers without contaminating the speaker embedding space well trained for base speakers. The synthesized audio samples are presented at  \url{https://prml-lab-speech-team.github.io/demo/GC-TTS}

\section{MODEL ARCHITECTURE}
\label{sec:model}

\subsection{Multi-Speaker TTS System}
\label{model_multi-TTS}
The model architecture of GC-TTS is rooted in Tacotron2\cite{shen2018natural:1} with the extension of a neural speaker encoder and a classification layer, as illustrated in Fig.~\ref{fig:network}. The speaker encoder takes a log mel-spectrogram of the target speaker as input and computes the corresponding speaker embedding which is then concatenated with the Tacotron2 encoder output that the decoder will attend to. We follow the structure in \cite{skerry2018towards:2} for the speaker encoder which comprises six convolution blocks followed by Gated Recurrent Unit (GRU) layer\cite{cho2014learning}, and a Fully Connected (FC) layer. Each convolution block has 2d convolutional layer, batch normalization\cite{ioffe2015batch} and relu activation\cite{nair2010rectified}. We also add a classification layer that takes a speaker embedding as an input and predicts the corresponding speaker label. By applying two geometric constraints, speaker embeddings of utterances from the same speaker cluster near the corresponding classification weight, and each classification weight of a novel speaker is separated from the classification weights of the other speakers.

\begin{table*}[!ht]
    \caption{Evaluation results. MOS-Nat. and MOS-Sim. refer to MOS for naturalness and similarity, respectively. Lower is better for MCD$_{13}$ and RMSE$_{f0}$, and higher is better for the other metrics. MOS results are presented with 95\% confidence intervals. $\emph{1m}$ and $\emph{5m}$ represent 1 and 5 minutes of adaptation data, respectively. Values in bold represent the best results for each metric.}
    \centering
    \begin{tabular}{c|c|c|c|c|c|c} \Xhline{3\arrayrulewidth}
        \textbf{Method} &\textbf{MOS-Nat.$(\uparrow)$} &\textbf{MOS-Sim.$(\uparrow)$} &\textbf{MCD$_{13} (\downarrow)$} &\textbf{RMSE$_{f0} (\downarrow)$} &\textbf{Cos.-Sim. $(\uparrow)$} &\textbf{Acc. $(\uparrow)$} \\ \hline     
            Ground Truth &$4.05\pm{0.03}$ &$4.29\pm{0.02}$ &- &- &- &$99.75\%$    \\
            Vocoded &$3.95\pm{0.04}$ &$4.19\pm{0.04}$ &$2.903$  &$35.153$ &$0.9723$ &$95.88\%$  \\
            Tacotron2 \cite{shen2018natural:1} &$3.84\pm{0.04}$ &$3.60\pm{0.04}$ &$4.869$ &$37.630$ &$0.8942$ &$92.63\%$   \\ \hline 
            Jia et al.\cite{jia2018transfer:7} &$3.83\pm{0.04}$ &$3.15\pm{0.05}$ &$5.766$ &$43.068$ &$0.7806$ &$36.75\%$   \\ 
            Cooper et al. \cite{cooper2020zero} &$\textbf{3.86}\pm{\textbf{0.04}}$ &$3.19\pm{0.03}$ &$5.818$  &$46.712$ &$0.7960$ &$32.63\%$    \\ \hline
            Fine-tuning $(<\emph{1m})$ &$3.72\pm{0.04}$ &$3.25\pm{0.05}$ &$5.484$ &$41.516$ &$0.8077$ &$65.25\%$   \\ 
            GC-TTS $(<\emph{1m})$ &$3.78\pm{0.04}$ & $3.39\pm{0.05}$ &$5.415$ &$38.811$ &$0.8211$ &$73.50\%$    \\
            Fine-tuning $(<\emph{5m})$ &$3.74\pm{0.04}$ &$3.34\pm{0.05}$ &$5.375$ &$40.278$ &$0.8173$ &$70.38\%$   \\ 
            GC-TTS $(<\emph{5m})$ &$3.81\pm{0.04}$ &$\textbf{3.45}\pm{\textbf{0.05}}$ &$\textbf{5.345}$ &$\textbf{37.858}$ &$\textbf{0.8354}$ &$\textbf{78.13\%}$  \\                
       \Xhline{3\arrayrulewidth}
    \end{tabular}
    \label{MOS}
\end{table*}

\subsection{Two-Stage Training Procedure}
\label{two-stage}
We use a two-stage training procedure. In training stage 1, GC-TTS is pre-trained for base speakers with a large amount of training data. Here, the speaker encoder is trained jointly with the Tacotron2's reconstruction loss and stop token prediction loss\cite{shen2018natural:1}. To train a classifier, we use the softmax layer followed by cross-entropy loss defined as follows:
\begin{equation}
    L_{cls}=-\frac{1}{M}\sum_{i=1}^{M} log\frac{exp({e_{i}^{T} } w^{y_i})}  {{ \sum_{j=1}^{n}exp(e_{i}^{T}w^{j})}}     
    \label{cross-entropy}
\end{equation}
where $M$ represents a batch size, $n$ refers to the number of training examples, $y_i$ represents the speaker label of the $i$th example, $e_i$ is the speaker embedding computed from the $i$th example, and $w$ represents the weights of classifier.

In training stage 2, we fine-tune the pre-trained network for novel speakers with a few amount of adaptation data. We first freeze the specific modules of the model trained for base speakers in training stage 1. The intuition for this is that the more trainable parameters it has, the more it is vulnerable to overfitting. Following \cite{moss2020boffin}, we freeze the weights of the Tacotron2 encoder and the attention module since they are already well trained across diverse texts and multiple speakers with various prosodies. Additionally, we freeze the weights of \emph{K} bottom convolution blocks in the speaker encoder. This is due to the fact that the bottom convolution blocks tend to learn \emph{general} features across various speakers\cite{yang2007reconstruction,yosinski2014transferable}. In practice, we find that a \emph{K} of 4 can lead to the most robust synthesis as we will validate it in Sec~\ref{ablation-experi}.

After freezing the parts of the network, we replicate the other modules to be fine-tuned and build a novel classifier\cite{lee1999integrated}. Based on this network, we employ two geometric constraints in training stage 2, which are described in Section~\ref{sec:loss}. The adaptation data of novel speakers is forwarded to the frozen parts of the network, the replicated parts of the network, and the novel classifier. Since the base modules aren't optimized during the adaptation, the generation quality for base speakers can be preserved even after the adaptation process. This is a simple but essential technique to preserve generation quality for base speakers.

\subsection{Vocoder}
To convert the predicted mel-spectrograms into a raw waveform\cite{yoon2020audio}, we use a Parallel WaveGAN\cite{yamamoto2020parallel} as our vocoder. With Wavenet\cite{oord2016wavenet}-based generator and multi-resolution spectrogram training objectives, it achieves high-quality waveform generation in real-time.

\begin{table*}[!t]
    \caption{The results of ablation study for two geometric constraints.}
    \centering
    \begin{tabular}{l|c|c|c|c|c|c} \Xhline{3\arrayrulewidth}
       
     \textbf{Setting} &\textbf{MOS-Nat. $(\uparrow)$} &\textbf{MOS-Sim. $(\uparrow)$} &\textbf{MCD$_{13} (\downarrow)$} &\textbf{RMSE$_{f0} (\downarrow)$} &\textbf{Cos.-Sim. $(\uparrow)$} &\textbf{Acc. $(\uparrow)$}     \\  \hline
            \textbf{Proposed} &$\textbf{3.76}\pm{\textbf{0.04}}$ &$\textbf{3.29}\pm{\textbf{0.05}}$ &$\textbf{5.415}$ &$\textbf{38.811}$ &$\textbf{0.8211}$ &$\textbf{73.50\%}$  \\ \hline
            $-\mathcal{L}_{WCEC}$ &$2.28\pm{0.04}$  &$2.65\pm{0.06}$ &$6.542$ &$41.851$ &$0.7024$ &$45.00\%$  \\ 
            $-\mathcal{L}_{AWS}$ &$3.49\pm{0.03}$ &$3.18\pm{0.04}$ &$5.468$ &$40.526$ &$0.7830$ &$66.25\%$  \\ 
            $-\mathcal{L}_{WCEC}-\mathcal{L}_{AWS}$ &$3.26\pm{0.03}$ &$3.01\pm{0.05}$ &$5.980$ &$42.517$ &$0.7547$ &$59.63\%$  \\
        \Xhline{3\arrayrulewidth}
    \end{tabular}
    \label{ablation_loss}
\end{table*}

\section{LOSS FUNCTIONS}
\label{sec:loss}
Critical to good generalization is to extract discriminative speaker embeddings of novel speakers from limited data. To achieve this, we apply three loss functions in training stage 2: cross-entropy loss and two geometric constraints proposed in \cite{jung2020few:11}. For the two geometric constraints, we first locate the speaker embedding space to a hypersphere by $l2$-normalizing the speaker embeddings and the classification weights. This eliminates information related to the magnitude and thus makes it possible to geometrically control the speaker embedding space by using only angular information, which is highly beneficial to few-shot learning settings\cite{jung2020few:11}. In addition to these loss functions, Tacotron2's reconstruction loss and stop token prediction loss are also used as in training stage 1.

\subsection{Weight-Centric Embedding Clustering (WCEC)}
If we train the network for novel speakers from scratch using a few data, speaker embeddings of novel speakers may not be well clustered. To address this issue, we employ Weight-Centric Embedding Clustering (WCEC) loss:
\begin{equation}
    \mathcal{L}_{WCEC}= \sum_{i=1}^{n_{N}}-log(cos\theta_{g(e^i),\widetilde{w}_{N}^{i}})
\end{equation}
where $n_{N}$ refers to the number of the novel speakers and $cos\theta_{g(e^i), \widetilde{w}_{N}^{i}}=g(e^i)^{T}\cdot(w_{N}^{i}/||w_{N}^{i}||)$ is the cosine similarity between $g(e^i)$ and the normalized classification weight of the $i$th novel speaker, denoted $\widetilde{w}_{N}^{i}$. The function $g(\cdot)$ is repre-sented as follows:
\begin{equation}
    g(e^i)=\frac{\bar{e^i}} {||\bar{e^i}||}    
\end{equation}
Here, $\bar{e^i}$ is the arithmetic mean of the normalized speaker embeddings of the $i$th novel speaker. We use $\bar{e^i}$ as the initial classification weight of the $i$th novel speaker, so that the cosine similarity can be initialized with a positive number and this WCEC loss can be defined.  

When used with cross-entropy loss, this WCEC loss enables the speaker embeddings of novel speakers to be clustered near the corresponding classification weight.

\subsection{Angular Weight Separation (AWS)} 
Since the WCEC loss has clustered the speaker embedd-ings of novel speakers near the corresponding classification weight, now we aim to separate each classification weight apart from the others. To this end, we first clarify the cosine similarity $u_{ij}$ between the classification weights:
\begin{equation}
    u_{ij}=
    \begin{cases}
        0, & \mbox{if }\widetilde{w}^i\equiv \widetilde{w}^j_N \\
        cos\theta_{\widetilde{w}^i,\widetilde{w}_N^j}, & \mbox{otherwise}
    \end{cases}
\end{equation}
Here, $\widetilde{w}^i$ is the $i$th column vector of $[\widetilde{W}_B\widetilde{W}_N]$, which is the normalized weight matrix of base and novel classifier, denoted $\widetilde{W}_B$ and $\widetilde{W}_N$, respectively. Our goal is to reduce $u_{ij}$ by margin $m$ to enlarge angular distance $\theta_{\widetilde{w}^i,\widetilde{w}_N^j}$ between the classification weights. This AWS loss is defined as follows:
\begin{equation}
    \mathcal{L}_{AWS} = \cfrac{\sum_{i,j}-log(-u_{ij}\cdot \mathbb{1}(u_{ij})+1) }
    { \sum_{i,j}\mathbb{1}(u_{ij}) }
\end{equation}
where $\mathbb{1}(u_{ij})=\begin{cases}1,&\mbox{if\enskip}\forall i,j,\enskip u_{ij}>m \\0,&\mbox{otherwise}\end{cases}$.

The logarithm and the factor of $1$ are used to align AWS loss with the scale of other loss functions\cite{jung2020few:11}. When each classification weight of a novel speaker is sufficiently separated from that of the other speakers in terms of the angular distance, we stop using the AWS loss. In all our experiments, we use a margin of $m = 0.5$ and all the losses have equal weights.

\section{EXPERIMENTS}
\subsection{Experimental Settings}
We conducted experiments on the VCTK dataset, which contains 108 English native speakers with various accents. The audio was downsampled from $48,000$Hz to $22,050$Hz and the corresponding transcripts were converted from character to phoneme sequence. We randomly chose 85 speakers for the base speakers and 16 speakers for the novel speakers. The remaining 7 speakers were used for the validation set. For each novel speaker, 10 and 50 samples were randomly sampled for the adaptation data (representing less than 1 and 5 minutes, respectively). We constructed the test set for novel speakers that do not appear in any training sets.

We followed the implementation details in the work of \cite{shen2018natural:1} with some modifications. In the training stage 1, we used $32$ of batch size, and $10^{-3}$ of the initial learning rate. The learning rate was decayed by a 0.5 factor in every $50$K step with the minimum learning rate of $10^{-5}$. In the training stage 2, we used $8$ of batch size and $10^{-4}$ of fixed learning rate. We utilized the ADAM optimizer\cite{kingma2014adam} with $\beta_1$ = 0.9, $\beta_2$ = 0.98, and $\epsilon$ = $10^{-9}$. 80 bands mel-spectrogram was transformed with 1024 of window size, 256 of hop size, and 1024 points of Fourier transform. 

\subsection{Evaluation Method}
Along with ground truth audio and the vocoded audio from the ground truth mel-spectrogram (Vocoded), we compared our GC-TTS against several baselines. Since they used various experimental setups, the results were from the most similar configuration to ours.
\begin{itemize}
    \item \textbf{Tacotron2 \cite{shen2018natural:1}} refers to multi-speaker Tacotron2 trained with hours of training data for novel speakers (about 30 minutes for each novel speaker). Since it was trained on a large amount of data, it can be regarded as our upper bound.
    \item \textbf{Jia et al.\cite{jia2018transfer:7}} transfers well-trained speaker embedding computed from the speaker verification model of \cite{wan2018generalized:18} into pre-trained multi-speaker Tacotron2. We followed the open-source implementation \cite{ge2e:impl} to implement the model of \cite{wan2018generalized:18}.
    \item \textbf{Cooper et al. \cite{cooper2020zero}} exploits speaker embedding com-puted from external speaker verification model of \cite{cai2018exploring}. We utilized the official implementation \cite{lde:impl} to extract speaker embeddings for multi-speaker Tacotron2.
    \item \textbf{Fine-tuning} is a strong comparable system that fine-tunes the whole model of Tacotron2 without using two geometric constraints and freezing weights in a similar scheme to \cite{arik2018neural:5, chen2018sample:6}. 
\end{itemize}

We measured the quality of synthesized speech on two scales. As a subjective test, we performed MOS tests. As an objective assessment, we computed MCD$_{13}$, RMSE$_{f0}$, cosine similarity, and speaker classification accuracy.
\begin{itemize}
    \item \textbf{MOS} stands for Mean-Opinion-Score. We performed MOS tests via Amazon MTurk, where at least 20 subjects were asked to rate the naturalness and speaker similarity on a score of 1-5 in 0.5 point increments. 
    \item \textbf{MCD$_{13}$} represents mel-cepstral distortion \cite{kubichek1993mel}. It compares mel-frequency cepstral coefficient between ground truth and synthesized audio signals.
    \item \textbf{RMSE$_{f0}$} computes $l2$ distance between fundamental frequencies of ground truth and that of synthesized audio\cite{hayashi2017investigation}.
    \item \textbf{Cosine Similarity} measures how similar the voice of the synthesized speech is to that of the target speaker. We extracted x-vectors\cite{snyder2018x} from generated and the actual speech, then calculated the cosine similarity between them.
    \item \textbf{Classification Accuracy} is another metric for evaluating speaker similarity between synthesized and the actual voice. We trained a \emph{test-only} speaker classification network using x-vectors\cite{snyder2018x} extracted from the actual speech of the novel speakers.
\end{itemize}

\begin{table}[!t]
        \caption{The results of ablation study for freezing modules. Enc and Attn refer to Tacotron2 encoder and attention module, respectively. SPK-ENC K denotes freezing K bottom convolution blocks in the speaker encoder.}
        \centering
        \begin{tabular}{l|c|c|c|c} \Xhline{3\arrayrulewidth}
           
         \textbf{Setting}  &\textbf{MCD$_{13}$} &\textbf{RMSE$_{f0}$} &\textbf{Cos.-Sim.} &\textbf{Acc.}  \\  \hline     
                Nothing  &$5.511$ &$41.676$   & $0.8092$ &$68.50\%$\\ 
                \emph{+Enc + Attn}  &$5.447$ &$39.483$   &$0.8105$  &$66.50\%$\\
                \emph{+SPK-ENC1} &$5.498$ &$40.536$ &$0.8101$ &$67.62\%$ \\ 
                \emph{+SPK-ENC2} &$5.465$ &$40.820$ &$0.8115$ &$68.00\%$ \\ 
                \emph{+SPK-ENC3} &$5.479$ &$\textbf{38.294}$ &$0.8119$ &$70.13\%$ \\
                \textbf{\emph{+SPK-ENC4}} &$\textbf{5.415}$ &$38.811$ &$\textbf{0.8211}$ &$\textbf{73.50\%}$ \\
                \emph{+SPK-ENC5} &$5.469$ &$39.156$ &$0.8174$ &$68.38\%$ \\
                \emph{+SPK-ENC6} &$5.436$ &$41.767$ &$0.8080$ &$66.00\%$ \\
    
            \Xhline{3\arrayrulewidth}
        \end{tabular}
        \label{ablation_frz}
\end{table}

\begin{figure*}[!t]
    \centering
    \scriptsize
    \subfigure[Trained with base speakers]{
    \centering
    \includegraphics[width=.55\columnwidth]{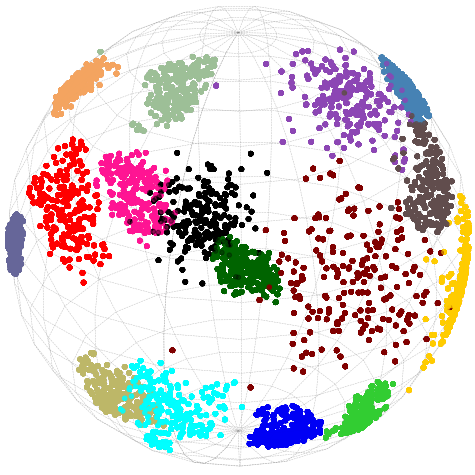}
    \label{tsne:base}
    }
    \subfigure[Fine-tuned w/o geometric constraints]{
    \includegraphics[width=.55\columnwidth]{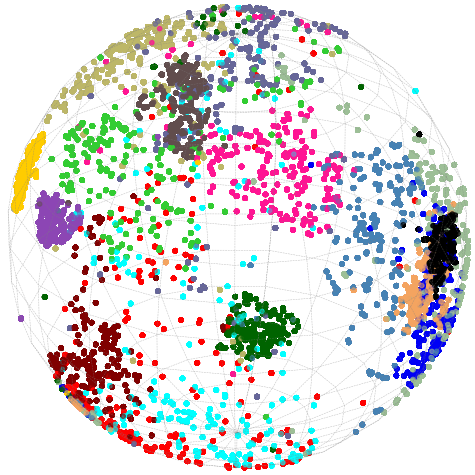}
    \label{tsne:finetune}
    }
    \subfigure[Fine-tuned w/ geometric constraints]{
    \includegraphics[width=.55\columnwidth]{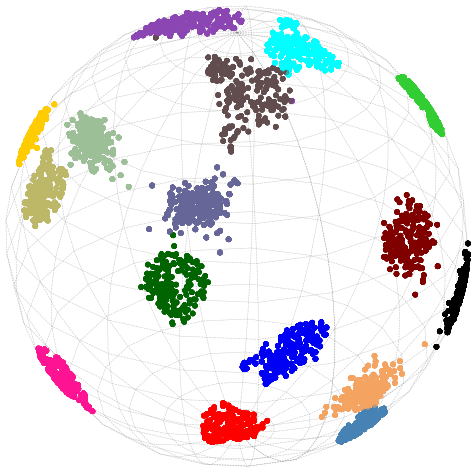}
    \label{tsne:proposed}
    }
    \subfigure{
    \includegraphics[height=5cm, width=1.3cm]{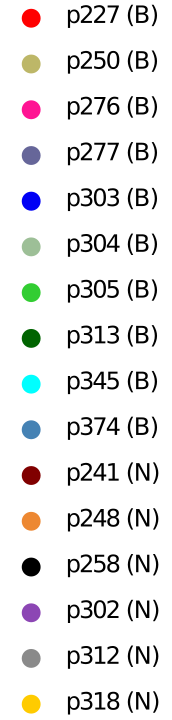}
    \label{tsne:label2}
    }

    \caption{(Best viewed in color)
    t-SNE visualization of speaker embedding space of base and novel speakers. The label of (B) refers to the base speakers and (N) represents the novel speakers. (a) Speaker embedding space computed from the model trained only with base speakers. (b) Speaker embedding space after the whole model is merely fine-tuned. (c) Speaker embedding space after the model is adapted to the novel speakers by the proposed method.}
\label{tsne}
\end{figure*}

\section{RESULTS AND ANALYSIS}
\label{sec:experiments}
\subsection{Subjective Evaluation}
To evaluate the adaptation performance of GC-TTS in terms of both naturalness and speaker similarity, we conducted the MOS test. For the MOS test, we randomly sampled $80$ generated speech from the novel speakers.

Table~\ref{MOS} shows the MOS results compared to our baselines. \cite{jia2018transfer:7} and \cite{cooper2020zero} achieved better naturalness than GC-TTS because their decoder followed well-trained base speakers; this is not the desired result since we intend to generate natural speech in the voice of novel speakers. In terms of the speaker similarity which is the main metric for the few-shot speaker adaptation problem \cite{chen2018sample:6}, GC-TTS achieved the best MOS result. This implies that the perceptual similarity of the speech generated by GC-TTS is better than the others.

\subsection{Objective Evaluation}
As objective assessments, we computed MCD$_{13}$\cite{kubichek1993mel}, RMSE$_{f0}$\cite{hayashi2017investigation}, cosine similarity, and speaker classification accuracy using randomly sampled 800 generated speech of the novel speakers.  

As demonstrated in Table~\ref{MOS}, GC-TTS shows the best results in every objective metric. Specifically, the speaker classification accuracy was highly improved, which demonstrates that the proposed model clones a much similar voice to the target speaker than the other models. In addition, when adapted with 50 samples (\emph{$<$ 5m}), GC-TTS achieves the highest score in RMSE$_{f0}$ with a gap of only $0.228$ compared to the Tacotron2 which can be regarded as our upper bound. 

\subsection{Ablation Study} 
\label{ablation-experi}
We conducted an ablation study of two geometric constraints, i.e. WCEC loss and AWS loss, to verify the effect of them. For brevity, we used 10 samples of each novel speaker as the adaptation data, and each system was trained until convergence.

The results are shown in Table~\ref{ablation_loss}, which demonstrates all the geometric constraints contribute to the adaptation performance in terms of both naturalness and speaker similarity. Especially, removing WCEC loss leads to inconsistent speaker embedding, and thus significantly degrades the generation performance.

Moreover, we performed another ablation study of freezing modules to validate the statement in Sec.~\ref{two-stage}. The results are shown in Table~\ref{ablation_frz}. We can get the most natural synthesis when we freeze Tacotron2 encoder, the attention module, and 4 of bottom convolution blocks. Whereas, freezing no modules show the worst result in MCD$_{13}$.

\subsection{Speaker Embedding Space}
We visualize the speaker embedding space using t-Stochatic Neighbor Embedding (t-SNE) visualization. From all 85 base speakers and 16 novel speakers, we randomly chose 12 base speakers and 6 novel speakers and computed their speaker embeddings using 200 utterances of each speaker.

Fig.~\ref{tsne} represents the speaker embedding space of the model fine-tuned with or without two geometric constraints. Fig.~\ref{tsne:base} represents the speaker embedding space computed from the model trained with only the base speakers. The speaker embeddings of the base speakers are well trained but those of the novel speakers are spread out or overlapped with those of base speakers. When the whole network is merely fined-tuned with 10 samples of each novel speaker, the speaker embeddings of the novel speakers somewhat cluster but it damages those of the base speakers as in Fig.~\ref{tsne:finetune}. From Fig.~\ref{tsne:proposed}, it is evident that GC-TTS not only constructs discriminative speaker embeddings of the novel speakers but also preserves the speaker embeddings of the base speakers using only 10 adaptation data from each novel speaker.

\begin{comment}
    \begin{table}[ht]
        \caption{Abalation study results1}
        \centering
        \begin{tabular}{l|c|c|c} \Xhline{3\arrayrulewidth}
           
         \textbf{Freeze Modules}& \textbf{MCD}$_{13}$ &\textbf{Cos-sim.} & \textbf{Acc.}  \\  \hline     
                Nothing &$5.498$   & $0.8174$ &$68.50\%$\\ 
                \emph{+Enc + Attn} &$5.447$   &$0.8105$  &$66.50\%$\\
                \emph{+SPK-ENC1} &$5.511$ &$0.8101$ &$67.62\%$ \\ 
                \emph{+SPK-ENC2} &$5.465$ &$0.8115$ &$68.00\%$ \\ 
                \emph{+SPK-ENC3} &$5.479$ &$0.8119$ &$70.13\%$ \\
                \emph{+SPK-ENC4} &$5.436$ &$\textbf{0.8211}$ &$\textbf{73.50\%}$ \\
                \emph{+SPK-ENC5} &$5.469$ &$0.8092$ &$68.38\%$ \\
                \emph{+SPK-ENC6} &$\textbf{5.416}$ &$0.8080$ &$66.00\%$ \\
    
            \Xhline{3\arrayrulewidth}
        \end{tabular}
        \label{Speaker Accuracy}
\end{table}
\end{comment}

\section{Conclusion and future works} 
\label{sec:conclusion}
We presented the few-shot speaker adaptive framework with two geometric constraints. The two geometric constraints enabled speaker embedding space to be controlled by only angular information, so that we could extract discriminative speaker embeddings of novel speakers from only a few data while preserving speaker embedding space of base speakers. From the experiments, we demonstrated that GC-TTS significantly improved the speaker similarity of generated speech over standard techniques. The geometric constraints are highly scalable in that they can be applied not only to image classification but also to speaker adaptation. For future work, we will apply the geometric constraints to other tasks such as emotional TTS or cross-lingual problems.

\section*{ACKNOWLEDGMENT}
The authors would like to thank H.-R. Noh at Netmarble Corp., Seoul, Korea, and H. Chung at Korea Univ., Seoul, Korea, for their valuable discussions and supports.
\bibliographystyle{IEEEtran} % We choose the "plain" reference style
\bibliography{mybib} % Entries are in the "refs.bib" file

\end{document}